\documentclass[twocolumn,twoside,10pt,superscriptaddress,prl]{revtex4}
\usepackage{amssymb}

\usepackage{amsmath}
\usepackage{graphicx}

\input{tcilatex}

\begin{document}

\title{Implementations of Nonadiabatic Geometric Quantum Computation using
NMR}
\author{Jiangfeng Du}
\email{djf@ustc.edu.cn}
\affiliation{Department of Modern Physics, University of Science and Technology of China,
Hefei, 230027, People's Republic of China.}

\author{Mingjun Shi}
\affiliation{Department of Modern Physics, University of Science and Technology of China,
Hefei, 230027, People's Republic of China.}

\author{Jihui Wu}
\affiliation{Laboratory of Structure of Biology, University of Science and Technology of
China, Hefei, 230027, People's Republic of China.}

\author{Xianyi Zhou}
\affiliation{Department of Modern Physics, University of Science and Technology of China,
Hefei, 230027, People's Republic of China.}

\author{Rongdian Han}
\affiliation{Department of Modern Physics, University of Science and Technology of China,
Hefei, 230027, People's Republic of China.}

\begin{abstract}
Recently, geometric phases, which is fault tolerate to certain errors
intrinsically due to its geometric property, are getting considerable
attention in quantum computing theoretically. So far, only one experiment
about adiabatic geometric gate with NMR through Berry phase has been
reported. However, there are two drawbacks in it. First, the adiabatic
condition of Berry phase makes such gate very slowly. Second, the extra
operation to eliminate the dynamic phase. As we know, geometric phase can
exist both adiabatic(Berry phase) and nonadiabatic(Aharonov-Anandan phase).
In this letter, we reports the first experimental realization of
nonadiabatic geometric gate with NMR through conditional-AA phase. In our
experiment the gates can be made faster and more easily, and the two
drawbacks mentioned above are removed.
\end{abstract}

\maketitle

Quantum computers can perform certain tasks much more efficiently than
classical Turing Machine\cite{Deutch}. It is well known that controlled
two-qubit gate, combined with single qubit operations, is a universal gate
for quantum computation\cite{Barenco}. This two-qubit gate preserves the
target qubit for the controlling qubit in certain state, say, $\left|
\uparrow \right\rangle $, and flips the target qubit for the controlling
qubit in the other state, $\left| \downarrow \right\rangle $. Originally
this has been achieved experimentally using dynamic method in different
physical systems\cite{ion,QED,nmr}.

On the other hand, central to the experimental realization of quantum
computer is the construction of fault-tolerant quantum logic gates\cite%
{steane}. In the quest for a low noise quantum computing device, geometric
phases\cite{Berry,AA}, which is fault tolerate to certain errors
intrinsically due to its geometric property, are getting considerable
attention in quantum computing theoretically\cite%
{jones,vedral,duan,Pellizarri,
Averin,Zanardi,Pachos,holonomic1,holonomic2,holonomic3}. Geometric logical
gates based on Berry phase\cite{Berry} restricted in adiabatic evolution
have been proposed\cite{vedral,duan}, and the first adiabatic geometric gate
was implemented in NMR\cite{jones}. However, the experiments rely on the
adiabatic operations. This is a bit impractical because the experimental
result is inexact unless the Hamiltonian changes extremely slowly in the
process. But in fact, everything has to be completed within the decoherence
time. Besides the adiabatic condition, both of the previous proposals
require extra operation to eliminate the dynamic phase. This extra operation
is unwanted for a fault tolerate gate because if we can not eliminate the
dynamic phase exactly, the fault tolerate property is weakened. For these
reasons, one is tempted to set up a new scheme which does not rely on the
adiabatic condition and which does not involve any accessory operation
performed to eliminate dynamic phase in the whole process.

Indeed, geometric phase does exist in a non-adiabatic process. It was shown
by Aharonov and Anandan that the geometric phase for a two-level system is
only dependent on the area enclosed by the loop on the Bloch sphere\cite{AA}%
. In the non-adiabatic case, the path of the state evolution is, in general,
different from the path of the parameters in the Hamiltonian. The external
field need not always follow the evolution path of the state, in contrast to
the adiabatic case. Thus, it is possible to let the external field be
perpendicular to the evolution path instantaneously so that there is no
dynamic phase involved in the whole process. Recently, several schemes were
proposed for nonadiabatic geometric gates in different systems\cite%
{wang,Wellard,wang2,wang1,li}, but there have been no corresponding
experimental demonstrations. Here we show the realization of nonadiabatic
two-qubit gate through conditional-AA phase of one two-level subsystem
(qubit) controlled by the state of another qubit.

\begin{figure}[tbp]
\includegraphics{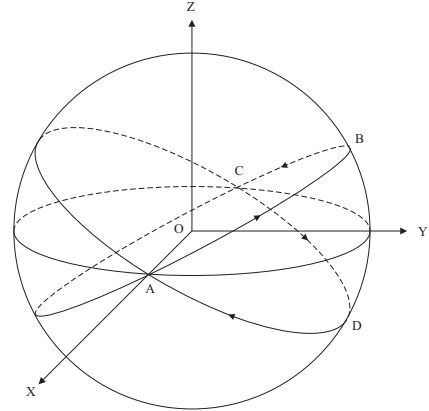}
\caption{The path evolution of qubit $a$
when qubit $b$ is in $\left| \uparrow \right\rangle $ state. The first pulse
(Fig.2) transformed the initial Hamiltonian $2\protect\pi JI_{z}^{a}$ to $H=2%
\protect\pi J\left( I_{z}^{a}\cos \protect\theta -I_{y}^{a}\sin \protect%
\theta \right) $ which creates an evolution path on the geodesic circle ABC.
After time $\protect\tau =1/\left( 2J\right) $, the second pulse changes the
Hamiltonian to the form $H=2\protect\pi J\left( -I_{z}^{a}\cos \protect%
\theta -I_{y}^{a}\sin \protect\theta \right) $ which creates an evolution
path on the geodesic circle CDA. Again after time $\protect\tau =1/\left(
2J\right) $, the third pulse restored the Hamiltonian to the initial form $2%
\protect\pi JI_{z}^{a}$. Therefore, qubit $a$ undergoes a cyclic evolution
through a slice circuit $C$ with angle $\protect\theta $\ in projective
Hilbert (density operator) space, and then the AA geometrical phase is
simply $\protect\beta \left( C\right) =m\ \Omega $, where $m=\pm \frac{1}{2}$
is the magnetic quantum number and $\Omega =4\protect\theta $ is the solid
angle subtended by the slice circuit.}
\label{Figure 1}
\end{figure}

Experimentally, the controlled two-qubit gate was implemented by the nuclear
spins of the $^{1}H$\ and $^{13}C$ atoms in a Carbon-13 labeled chloroform
molecule, the single $^{1}H$\ nucleus was used as target qubit $a$,\ while
the $^{13}C$\ nucleus was used as controlled qubit $b$, $\left| \uparrow
\right\rangle \left( \left| \downarrow \right\rangle \right) $ describes the
spin state aligned with (against) an externally applied, strong static
magnetic field $B_{0}$ in the $\widehat{z}$ direction. The reduced
Hamiltonian for this two-spin system is, to an excellent approximation,
given by $H=\omega _{a}I_{z}^{a}+\omega _{b}I_{z}^{b}+2\pi
JI_{z}^{a}I_{z}^{b}$, where the first two terms describe the free precession
of spin $a\left( ^{1}H\right) $ and $b\left( ^{13}C\right) $ about $B_{0}$
with frequencies $\omega _{a}/2\pi \approx 500Mhz$ and $\omega _{b}/2\pi
\approx 125Mhz$. $I_{z}^{a}$ ($I_{z}^{b})$ are the angular momentum operator
in the $\widehat{z}$ direction for $a$($b$). The third term of the
Hamiltonian describes a scalar spin-spin coupling of the two spins with $%
J=214.9Hz$. The spin-spin relaxation times are $0.3s$ for carbon and $0.4s$
for proton, respectively. Realistically, the state of this system is a
thermally equilibrium one, while pure $00$ state must be prepared for most
quantum computing tasks. To solve this problem, a variety of techniques
exist to extract from this thermal state just the signal from the molecules
in the $00$ state\cite{spacial-pure,time-pure, logic-pure}, i.e. all the
spins aligned in +z direction. Here we adopt the method of ``spatial
averaging'' to create this effective pure $00$ state\cite{spacial-pure}. The
initial state of spin $a$ is $\left| +\right\rangle =\frac{1}{\sqrt{2}}%
\left( \left| \uparrow \right\rangle +\left| \downarrow \right\rangle
\right) $, the start point of the path evolution, which is prepared by a 90$%
^{0}$ pulse oscillating at frequency $\omega _{a}$ along y-axes. This
resonance pulse rotate the pure state of qubit $a$ from +z-axes to +x-axes.

\begin{figure}[tbp]
\includegraphics{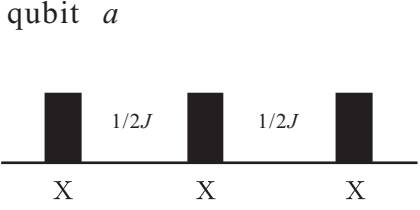}
\caption{Pulse sequence used to
demonstrate controlled-AA phase of the state of qubit $a$ or controlled
two-qubit gate. The black boxes are pulses oscillating at frequency $\protect%
\omega _{a}^{^{\prime }}=$ $\protect\omega _{a}-\protect\pi J$, the flip
angles are $-\protect\theta $, $2\protect\theta -\protect\pi $ and $\protect%
\pi -\protect\theta $ from left to right which can be realized by choosing
the different pulse duration $t$ and pulse power $P$. Here $\protect\theta $
was selected from $0$ to $\protect\pi $ by $\protect\theta =\frac{n\cdot 
\protect\pi }{16}$, $n=\left\{ 0,1,\cdots ,16\right\} $. All pulses that
oscillated at frequency $\protect\omega _{a}-\protect\pi J$ are hard pulses,
each pulse duration is $5us$. Delay times are $\protect\tau =1/2J$ between
pair of pulses. Then the time of nonadiabatic controlled two-qubit gate is
about $4.8ms$. Note this gate time do not depend on the value of AA phase.}
\label{Figure 2}
\end{figure}

To produce controlled-AA phase of the state, we change the oscillating
frequency $\omega _{a}^{^{\prime }}$ of qubit\ $a$\ from $\omega _{a}$ to $%
\omega _{a}-\pi J$. Now the Hamiltonian of qubit $a$ can be written as $%
H_{a}^{rot}=\left( \omega _{a}-\omega _{a}^{^{\prime }}\pm \pi J\right)
I_{z}^{a}=\left( \pi J\pm \pi J\right) I_{z}^{a}$, which describes the qubit
rotating around +z-axis in rotational frame with angular velocity $\omega
_{a}^{^{\prime }}=\omega _{a}-\pi J$. That is, when qubit $b$ is in the
state $\left| \uparrow \right\rangle $, $H_{a}^{rot}=2\pi JI_{z}^{a}$; when
qubit $b$ in $\left| \downarrow \right\rangle $, $H_{a}^{rot}=0$.
Consequently, different states of qubit $b$ correspond to different
frequencies of qubit $a$ in rotational frame. It is just this property that
we will use to realize `controlled-AA phase shift'. In Fig.1, we show the
cyclic evolution of $^{1}H$\ nucleus (qubit $a$) on the Bloch sphere when
the $^{13}C$ nucleus (qubit $b$) is in the state $\left| \uparrow
\right\rangle $. The pulse sequences to realize this nonadiabatic cyclic
evolution was shown in Fig.2. As we know, in toggling frame, the track of
Hamiltonian is defined by $H=H\left( 0\right) +H1(t)$, where $H\left(
0\right) $ is reduced Hamiltonian of the qubit $a$ $\left( \left( \pi J\pm
\pi J\right) I_{z}^{a}\right) $, and $H1(t)$ is Hamiltonian of RF pulses.
Therefore, by applying the pulse with angle $\theta $\ along x-axes(first
black box in Fig.2), the Hamiltonian is changed to $H=2\pi J\left(
I_{z}^{a}\cos \theta -I_{y}^{a}\sin \theta \right) $ in toggling frame.
Subject to this Hamiltonian, qubit $a$ undergoes procession from initial
state $\frac{1}{\sqrt{2}}\left( \left| \uparrow \right\rangle +\left|
\downarrow \right\rangle \right) $. After time $\tau =1/\left( 2J\right) $,
the initial state becomes $\left| \varphi \left( \tau \right) \right\rangle
_{a}=\frac{e^{-i\left( \pi /2+\theta \right) }}{\sqrt{2}}\left( \left|
\uparrow \right\rangle -\left| \downarrow \right\rangle \right) $. Seen from
the resulting state $\left| \varphi \left( \tau \right) \right\rangle _{a}$,
the direction of spin $a$ is along --x-axis, besides a global phase factor.
Now by another pulse with angle $2\theta -\pi $ the Hamiltonian of qubit $a$
has the form $H=2\pi J\left( -I_{z}^{a}\cos \theta -I_{y}^{a}\sin \theta
\right) $. Again, subject to this Hamiltonian, qubit $a$ undergoes
procession from state $\left| \varphi \left( \tau \right) \right\rangle _{a}$%
. After time $\tau =1/\left( 2J\right) $, the state becomes $\left| \varphi
\left( 2\tau \right) \right\rangle _{a}=\frac{e^{-2i\theta }}{\sqrt{2}}%
\left( \left| \uparrow \right\rangle +\left| \downarrow \right\rangle
\right) $. The direction of spin $a$ returns to +x-axis and a phase $%
-2\theta $ appears. Finally we restore Hamiltonian of qubit $a$ to the
initial form $2\pi JI_{z}^{a}$ by the last pulse with angle $\pi -\theta $.
Note that this pulse has no influence on the final state of qubit $a$, i.e. $%
\left| \varphi \left( 2\tau \right) \right\rangle _{a}$.

From the whole process described above, the Hamiltonian of qubit $a$ has
experienced a cyclic evolution because of the cyclic property of the pulse
sequence. Correspondingly, an evolution path of ABCDA on the Bloch sphere is
produced for qubit $a$, that is $\frac{1}{\sqrt{2}}\left( \left| \uparrow
\right\rangle +\left| \downarrow \right\rangle \right) \rightharpoonup \frac{%
e^{-i\left( \pi /2+\theta \right) }}{\sqrt{2}}\left( \left| \uparrow
\right\rangle -\left| \downarrow \right\rangle \right) \rightharpoonup \frac{%
e^{-2i\theta }}{\sqrt{2}}\left( \left| \uparrow \right\rangle +\left|
\downarrow \right\rangle \right) $. Obviously the angle $\theta $ is the
geometric parameter describing the behavior of the Hamiltonian evolution.
This geometric parameter also describes the phase difference between the
initial and final state. Note that the dynamic phases appearing in two steps
of procession cancel out each other. The geometric AA phase is equal to $%
\beta \left( C\right) =-2\theta $. So far we have considered the case in
which the state of qubit $b$ is $\left| \uparrow \right\rangle $. If qubit $%
b $ is in $\left| \downarrow \right\rangle $, the Hamiltonian of qubit $a$
is zero and nothing will happen to it. In other words, qubit $a$ will
preserve itself in this case. Therefore, the time evolution operator of the
pulse sequence (Fig.2) has the property $U\left( 2\tau \right) \left| \pm
\right\rangle _{a}\left| \uparrow \right\rangle _{b}=e^{\pm i\beta \left(
C\right) }\left| \pm \right\rangle _{a}\left| \uparrow \right\rangle _{b}$
and $U\left( 2\tau \right) \left| \pm \right\rangle _{a}\left| \downarrow
\right\rangle _{b}=\left| \pm \right\rangle _{a}\left| \downarrow
\right\rangle _{b}$, where $\left| \pm \right\rangle $ corresponds to point
A and C, respectively, in the Bloch sphere. Hence we can regard qubit $b$ as
a controlling qubit and qubit $a$ as the target qubit; controlled-AA phase
of qubit $a$ is produced depended on the state of qubit $b$. In the basis of 
$\left| \uparrow \right\rangle $ and$\ \left| \downarrow \right\rangle $,
the unitary operator that describes this circle evolutin is%
\[
\left( 
\begin{array}{cc}
\begin{array}{cc}
\cos \left( \beta \left( C\right) \right) & i\sin \left( \beta \left(
C\right) \right) \\ 
i\sin \left( \beta \left( C\right) \right) & \cos \left( \beta \left(
C\right) \right)%
\end{array}
& 
\begin{array}{cc}
0 & 0 \\ 
0 & 0%
\end{array}
\\ 
\begin{array}{cc}
0 & 0 \\ 
0 & 0%
\end{array}
& 
\begin{array}{cc}
1 & 0 \\ 
0 & 1%
\end{array}%
\end{array}%
\right) . 
\]%
In particular, when $\left| \beta \left( C\right) \right| =\pi /2$, this
gate is just the C-not gate.

In order to measure the overall AA phase $\beta \left( C\right) $ of qubit $%
a $ we also apply a 90$^{0}$ pulse, oscillating at frequency $\omega _{a}$
along y-axes, to transform qubit $b$ in a coherent superposition of states $%
\left| \psi \left( 0\right) \right\rangle _{b}=\frac{1}{\sqrt{2}}\left(
\left| \uparrow \right\rangle +\left| \downarrow \right\rangle \right) _{b}$
before the cyclic evolution. Therefore after this cyclic evolution the final
state is $\frac{1}{\sqrt{2}}\left| +\right\rangle _{a}\left( \left| \uparrow
\right\rangle +\left| \downarrow \right\rangle \right) _{b}\rightharpoonup 
\frac{1}{\sqrt{2}}\left| +\right\rangle _{a}\left( \left| \uparrow
\right\rangle +e^{-i\beta \left( C\right) }\left| \downarrow \right\rangle
\right) _{b}$. Here, the unobservable AA phase $\beta \left( C\right)
=2\theta =\frac{1}{2}\Omega $ of qubit $a$ transfers to inner phase of qubit 
$b$ (13C in our experiments) which can be observed with NMR. To do so, we
use the signal of initial $\frac{1}{\sqrt{2}}\left( \left| \uparrow
\right\rangle +\left| \downarrow \right\rangle \right) _{b}$ state as a
reference one, compared with the signal of the final $\frac{1}{\sqrt{2}}%
\left( \left| \uparrow \right\rangle +e^{-i\beta \left( C\right) }\left|
\downarrow \right\rangle \right) _{b}$ state; both of them are in-phase
doublet but out of phase by $\beta \left( C\right) $. Experimentally, we
apply additional phasing factor $\beta \left( C\right) $ to obtain
absorptive lineshape after Fourier transformation.

All experiments are performed at room temperature and pressure on Bruker
Avance DMX-500 spectrometer in Laboratory of Structure Biology, University
of Science and Technology of China. The experimental results are shown in
Fig.3.

\begin{figure}[tbp]
\includegraphics{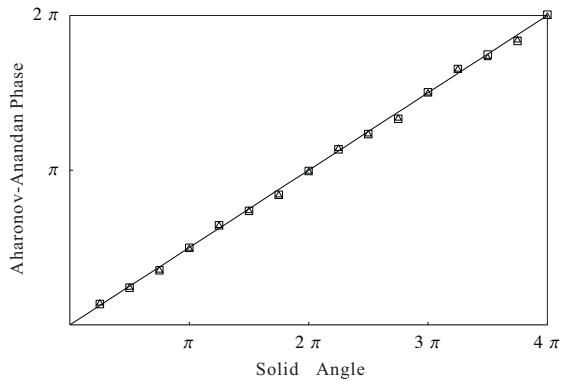}
\caption{Experimental values for the
controlled-AA phase $\protect\beta \left( C\right) $ as a function of solid
angle $\Omega =4\protect\theta $. Experimental points are shown as small
squares; theoretical values are shown as smooth curves. We can see the
experiment result fit the theory quite well; the remaining errors could be
due to phasing process which is influenced by machine noise and non-ideal
baseline. Besides, pulse imperfect and relaxation also have contributions.}
\label{Figure 3}
\end{figure}

Therefore, we have observed
controlled-AA phase or implemented a nonadiabatic two-qubit gate. Note that
though the observation of AA phase has been done in NMR with a three-level
system\cite{suter}, it cannot be used to implement universal two-qubit gate.

Our experiment resolves two drawbacks of the adiabatic geometric
computation, namely the slow evolution and the need of refocusing to
eliminate the dynamical phases. Let us now compare the gate time of this
nonadiabatic geometric gate to that of adiabatic geometric gate and dynamic
gate. Since the gate time is limited directly by the strength of coupling
constant $J$ of the sample. Two experiments we selected to compare used the
same sample as ours\cite{jones,nmr}, that is, Carbon-13 labelled chloroform
sample. In our experiment, it took about 4.8ms to realize the gate, slightly
longer than the time used to realize dynamic two-qubit gate (about 2.4ms)%
\cite{nmr}, yet much shorter than the time it took to realize the adiabatic
geometric two--qubit gate (about 120ms)\cite{jones}. As the adiabatic
geometric gate operates for a significantly longer time, it is much more
severely affected by decoherence. This has serious implications for the
physical realization of adiabatic geometric quantum computation. On the
other hand, since the state is always perpendicular to the effective
magnetic field, there is no dynamical phase accumulation during the
evolution, hence the resulted phase factor after cyclic evolution was pure
geometric phase. Although this nonadiabatic geometric gate is experimentally
realized in the NMR system, the basic idea is general, and could be applied
in other physical systems. We believe our experiment has led the idea of
geometric quantum computation much more practical than before.

We thank Y.D. Zhang, Z.B. Chen and S.Massar for discussion. This work was
supported by NSFC and CAS. Part of the ideas were originated while J.-F. Du
was visiting Service de Physique Theorique, Universite Libre de Bruxelles,
Bruxelles. S. Massar and N. Cerf are gratefully acknowledged for their
invitation and hospitality. J.-F. Du thanks X.-W. Zhu for the loan of the
chloroform sample.

\end{document}